# Judd-Ofelt analysis of holmium-doped alumino-silicate optical glass prepared by MCVD combined with nanoparticle doping

Petr Vařák[1,2], Michal Kamrádek[2], Pavla Nekvindová[1], Jan Hrabovský[3], Pavel Peterka[2]

[1]*Department of Inorganic Chemistry, University of Chemistry and Technology, Technická 5, 166 28, Prague, Czech Republic*

[2]*Institute of Photonics and Electronics of the Czech Academy of Sciences, Chaberská 1014/57, 182 00, Prague, Czech Republic*

[3]*Charles University, Faculty of Mathematics and Physics, Ke Karlovu 5, Prague 2, Prague, 121 16, Czech Republic*

varakp@vscht.cz

**Abstract**

We present a detailed Judd-Ofelt (JO) analysis of $Ho^{3+}$-doped alumino-silicate optical fiber preforms in a wide range of compositions and prepared by both standard solution doping method as well as the more advanced nanoparticle doping. The absorption spectra were measured, the absorption cross sections were calculated for each observed transition and the Judd-Ofelt analysis was conducted. The preforms exhibited the values of JO parameters of $\Omega_2$ = (7.2 – 10.6) $10^{-20}$ $cm^2$, depending on the $Al_2O_3$ content, and $\Omega_4$ and $\Omega_6$ around 2.5 and 1.2 $10^{-20}$ $cm^2$, respectively. The radiative transition parameters, such as transition probabilities, branching ratios and radiative lifetimes, were calculated. The radiative lifetime of the $^5I_7 \rightarrow ^5I_8$ transition, in the 16.1 – 17.2 ms range, combined with a measured value in the 0.9 – 1.9 ms range, was used to calculate the quantum efficiency, which was in the 5 – 11 % range. The preforms prepared by the nanoparticle doping containing high $Al_2O_3$ contents above 8 mol. % exhibited superior values of measured lifetime (above 1.5 ms) and quantum efficiency (above 8 %) compared to the standard solution-doped samples. To the best of our knowledge, this work is the first comprehensive report on the JO analysis of $Ho^{3+}$-doped alumino-silicate glass, the calculated parameters may be used in the various calculations, simulations and modelling of holmium-doped fiber lasers and other devices.

*Keywords*: Holmium, MCVD, Judd-Ofelt analysis, Absorption, Nanoparticle doping

## 1. Introduction

The trivalent ions of rare-earth (RE) metals from the $6^{th}$ period of the periodic table are well renowned for their sharp and intense absorption and emission peaks in the UV, visible and infrared regions, stemming from the transitions between various energy levels of the electrons contained within the *4f* inner shells [1]. The RE-doped materials thus find applications as the active elements in various types of lasers and amplifiers [2], as well as other devices, such as scintillators, i.e., detectors of high-energy radiation [3]. In the case of lasers and amplifiers based on optical fibers, the silica-based glass remains the leading material for the fabrication of the active optical fibers, due to its superior material properties, such as thermal, mechanical and chemical stability, as well as low cost and reliable, well-developed technology. However, pure silica glass represents an unsuitable environment for RE ions due to low RE solubility, causing concentration quenching, and high phonon energy, leading to multiphonon relaxation, i.e., thermal losses [4]. These issues may be at least partially overcome by the inclusion of $Al_2O_3$ in the glass matrix, and RE-doped alumino-silicate glass thus remains the primary option for the active element in fiber lasers or amplifiers [5]. These include devices based on $Yb^{3+}$ (emitting around 1 µm), $Er^{3+}$ ion (emitting around 1.5 µm), or $Tm^{3+}$ and $Ho^{3+}$ (both emitting around 2 µm). Specifically, holmium-doped fiber lasers (HFDL) recently attracted growing interest in development of high-power lasers for LIDARs or free-space communications owing to their emission in the 2.1 – 2.2 µm range, which corresponds to favorable atmospheric transmission [6–9].

However, the precise and effective design of new lasers and devices requires a thorough understanding of the material, in terms of both structural and spectroscopic properties. The answer to both questions may be provided by the Judd-Ofelt (JO) analysis. Introduced in 1962 by B. R. Judd [10] and G. S. Ofelt [11], the JO theory describes the intensities and probabilities of the *f-f* transitions of trivalent RE ions and more than 60 years later remains the cornerstone of RE material research [12,13]. The analysis allows us to use the measured absorption data to calculate quantities called line strengths or oscillator strengths of each observed transition and subsequently extract the three phenomenological JO parameters $\Omega_{2,4,6}$. These parameters have no direct physical meaning, but over the years, various correlations were drawn between the values of the parameters and the structural characteristics of the material, such as the symmetry of the RE ion sites, the covalency of the RE ion bonding, or the rigidity of the matrix [14].

More importantly, the parameters are used to calculate the theoretical line strengths, transition probabilities, branching ratios and radiative lifetimes of each individual transition, which may be in turn used to calculate, e.g., quantum efficiency or emission cross section. The theoretical emission parameters are also necessary as input in various theoretical tasks, such as calculations of energy transfer coefficients [15,16] or laser simulations [8,17]. Therefore, the precise JO analysis and the knowledge of these parameters is crucial for reliable material design and the development of novel lasers and amplifiers [4].

Despite the ubiquity of holmium-doped alumino-silicate glass in photonics, the JO analysis was only reported a handful of times, in all cases for alumino-silicate glass prepared by the sol-gel method. Moreover, the findings in literature are often conflicting and inconsistent, likely owing to significant differences in methodology, the crucial impact of which was previously described in [4,18]. Wang et al. [19] performed absorption measurements across the entire visible and near-infrared (NIR) spectral range and obtained $\Omega_{2,4,6}$ values of about 8.83, 2.45 and 1.06 ($10^{-20}$ $cm^2$ in all cases). Contrary, other works [20–22] presented similar samples in terms of composition and preparation, but using the absorption measurements only across the visible spectral range. The observed values of $\Omega_2$ in these studies are significantly lowered in the range 2 – 5 $10^{-20}$ $cm^2$ range. A similar methodology-dependable difference in the calculation of JO parameters was previously observed and described for $Er^{3+}$, where the selection of absorption bands and the omission of specific transitions lead to a significant distortion of results [4]. This observation led to the introduction of Combinatorial Judd-Ofelt analysis [18], but such a study is so far missing for $Ho^{3+}$ ion. As a result of the lack of reliable data, studies dealing with holmium-doped optical fiber and laser development often have to rely on outdated data, often from different glass systems [15,17], e.g., lithium-sodium-strontium-alumino silicate [23].

The recent progress in the HDFL research requires the clarification of this issue in order to obtain reliable numerical values of the radiative emission parameters and emission cross sections. In this work, we present the spectroscopic and JO analysis of holmium-doped alumino-silicate glass preforms prepared by the Modified Chemical Vapor Deposition (MCVD) technique combined with solution or nanoparticle doping method, same preforms which later serve as a precursor for drawing of laser-active optical fibers. The results of the JO analysis are reported, the quality of the prepared materials and their spectroscopic properties are discussed. The main result is the set of the emission parameters, i.e., transition probabilities, branching ratios and radiative lifetimes, for holmium-doped alumino-silicate glass in a wide range of compositions.

## 2. Experimental

### *2.1. Sample fabrication*

Optical fiber preforms doped with $Ho^{3+}$ ions and $Al_2O_3$ were prepared by the MCVD method combined with various types of doping based on the use of different types of solutions or dispersions (see lower). In all samples, a porous silica layer was deposited onto the inner wall of a pure silica tube (F300, Heraeus), soaked with an ethanolic solution or nanoparticle dispersion, and dried. The doped tubes were sintered under chlorine atmosphere at temperatures 1200–1800 °C and collapsed into preforms at temperatures up to 2100 °C. Two different types of doping were employed based on different type of doping solution or dispersion:

1. Traditional solution doping [24], using solution of $AlCl_3$ and $HoCl_3$ in ethanol. Samples marked as „SD“.
2. Nanoparticle doping [25,26], i.e., using ethanolic dispersion of commercial $Al_2O_3$ nanoparticles and dissolved $HoCl_3$. These samples can be further distinguished into two groups: a) low $Al_2O_3$ content up to 4 mol. %, comparable to solution-doped fibers; these samples are marked as L-NP, and b) higher $Al_2O_3$ content in 8 – 10 mol. % range, samples marked H-NP.

A more detailed information about the preform fabrication using both methods can be found in [26,27]. After a measurement of the refractive index profiles (RIPs) of the preforms, samples with approx. 5 mm thickness and 9 mm diameter were cut for further analysis. The samples were polished to optical quality using diamond polishing disks and CEROX suspension.

### 2.2. *Sample characterization*

The real composition of the fabricated preforms was measured by a Jeol JXA-8230 electron probe microanalyzer (EPMA) with the estimated measurement error of about 2 rel. %. The ion volume concentration was calculated using the glass density of 2.25 g $cm^{-3}$ for the samples containing up to 4 mol. %, and 2.30 g $cm^{-3}$ for the samples containing 8 – 10 mol. %, the densities were assumed based on [28]. The refractive index profiles of the preforms were measured using a Photon Kinetics A2600 refractive index profiler. The dispersion of the preforms was approximated using the established dispersion of pure silica glass expressed by Sellmeier equation [29] and adding the measured value of the refractive index for the preform. The approximated curve was re-fitted by Sellmeier function of the first order, equation (1).

$$n^2(\lambda) = A_S + \frac{B_S\,\lambda^2}{\lambda^2 - C_S} \tag{1}$$

The parameters $A_S$, $B_S$, and $C_S$ were obtained. The fluorescence decay curves of the 2 µm emission ($^5I_7 \rightarrow {}^5I_8$ transition) were measured using identical setup and procedure as described in [30] and the experimental fluorescence lifetime was obtained by single exponential fit. The absorption spectra of the preforms were measured using PerkinElmer Lambda 1050+ UV/Vis/NIR spectrophotometer in the 200 – 2250 nm range with 0.2 nm step, using photomultiplier (PMT) detector in 200 – 800 nm range, with 0.4 s integration time and 1 nm slit, and InGaAs detector in 800 – 2250 nm range, with 1.0 s integration time and 2 nm slit. A mask was applied on the samples to ensure transmission only through the core. The absorption coefficient, α, and absorption cross section, $\sigma_{ABS}$, were calculated from the measured transmittance, T, using equations (2) and (3).

$$\alpha(\lambda) = \frac{\ln 10}{L} \cdot \log_{10} \frac{1}{T(\lambda)}\ \ cm^{-1} \tag{2}$$

$$\sigma_{ABS}(\lambda) = \frac{\alpha(\lambda)}{N}\ cm^2 \tag{3}$$

Where *L* is sample thickness (cm) and *N* is the $Ho^{3+}$ ion volume concentration ($cm^{-3}$). The spectra were processed in OriginPro 2018b; the background absorption baseline was created by manual input and subtracted from data, the procedure was previously described and demonstrated in [31]. The absorption cross sections of the ground state absorption bands were integrated.

The JO analysis was calculated using the on-line software platform LOMS.cz (www.LOMS.cz). The entire procedure, including necessary equations, is described in [4] and [18]. The calculation was performed using ASCII input files containing all necessary parameters and values; the input files of all samples can be found in [32].

## 3. Results

### *3.1. General characterization*

The composition of the preform samples was determined by the EPMA method. For demonstration, the measured concentrations of $Al_2O_3$ and $Ho_2O_3$ at various points across the core of H-NP-5 preform sample are shown in Figure 1, forming homogenous concentration profiles, which nicely overlap with the refractive index profile. Other investigated preforms exhibited equivalent concentration and refractive index profiles.

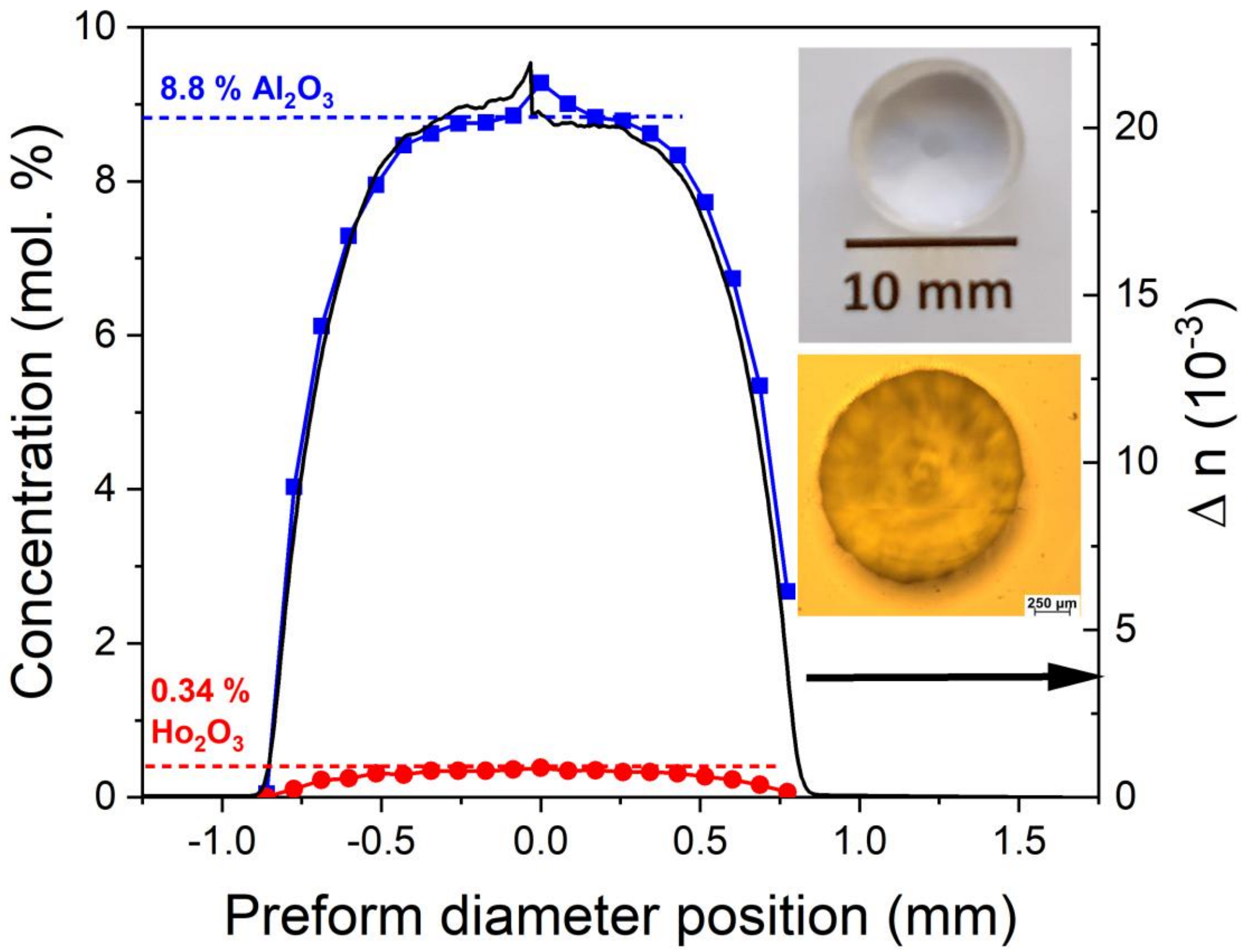

*Figure 1 – Concentration and refractive index profile of a representative preform H-NP-5, inset shows photograph of the preform, as well as of the core taken by optical microscope (please note that the final image of the core was made by splicing two original image files due to instrumental limitations, hence the color discontinuity edge inside the core).*

The samples prepared for this study and their composition are summarized in Table 1. In total, 9 preforms were prepared: 1 prepared by traditional solution doping containing approx. 3.5 mol. % Al2O3, a typical level for this traditional fabrication method [26]; 3 by the nanoparticle doping with comparable $Al_2O_3$ content up to 4 mol. %; and finally, the nanoparticle doping was employed to prepare 5 highly doped preforms containing 8 – 10 mol. % $Al_2O_3$. The concentration of $Ho^{3+}$ ions ranged between 1500 – 7000 ppm, covering the range typically found in fiber lasers [27].

*Table 1 – Estimated composition via EPMA of the preform samples*

| | $Al_2O_3$ (mol. %) | $Ho_2O_3$ (mol. %) | $Ho^{3+}$ ($10^{25}$ cm$^{-3}$) | Al/Ho |
|---|---|---|---|---|
| *Solution doping* | | | | |
| SD | 3.54 | 0.18 | 7.6 | 20 |
| *NP doping ($Al_2O_3$ < 4 mol. %)* | | | | |
| L-NP-1 | 2.90 | 0.11 | 4.8 | 26 |
| L-NP-2 | 3.90 | 0.16 | 7.0 | 24 |
| L-NP-3 | 2.70 | 0.31 | 13.6 | 9 |
| *NP doping ($Al_2O_3$ > 8 mol. %)* | | | | |
| H-NP-1 | 8.80 | 0.08 | 3.3 | 114 |
| H-NP-2 | 10.0 | 0.10 | 4.1 | 71 |
| H-NP-3 | 9.36 | 0.17 | 7.4 | 54 |
| H-NP-4 | 9.98 | 0.31 | 13.4 | 39 |
| H-NP-5 | 8.80 | 0.34 | 14.6 | 26 |

### *3.2. Absorption cross section*

The absorption cross section obtained according to experimental section is depicted in detail for the H-NP-5 preform in Figure 2. Numerous absorption bands, corresponding to transition from the $Ho^{3+}$:$^5I_8$ ground state to various excited states are marked in the spectrum. The positions (mean wavelengths) of these bands were nearly identical for all preforms, the exact values of mean wavelengths can be found in the LOMS software input files at [32]. Of special note is the absorption band corresponding to the $^5I_8 \rightarrow ^5I_5$ transition, located around 900 nm; this band was very weak and observable only in the H-NP-4 and H-NP-5 preforms. Similarly, the $^5I_8 \rightarrow ^5I_4$ absorption band is theoretically expected at approximately 750 nm, however, it has only rarely been observed in previously published studies and was not detected in the present work, even in the sample with the highest dopant concentration. The transition has very low reduced squared matrix-element values and is therefore intrinsically weak.

The absorption cross section data of all measured preforms are summarized in Figure 3. Several things may be noted. First, the absorption cross section values of the $^5I_8 \rightarrow ^5I_7$ (1800 – 2200 nm) and $^5I_8 \rightarrow ^5I_6$ (1100 – 1200 nm) transitions are nearly constant and independent on the Al or Ho concentration, amounting to $\sigma_{ABS}$(1940 nm) ≈ 3.8 $10^{-25}$ $m^2$ and $\sigma_{ABS}$(1150 nm) ≈ 1.5 $10^{-25}$ $m^2$, with only minor deviations. These values are in a reasonable agreement with literature, where the values of $\sigma_{ABS}$(1940 nm) were reported as 3.0 $10^{-25}$ $m^2$ for optical fiber [33] and 4.4 $10^{-25}$ $m^2$ for bulk sample [19]. Moreover, these values are in excellent agreement with our recent measurements on a set of optical fibers, where values of 3.5 $10^{-25}$ $m^2$ and 1.5 $10^{-25}$ $m^2$ were reported for 1940 and 1150 nm, resp. [34,35]; it must be noted these values represent a lower bound estimate owing to the intricacies of calculating absorption cross section in optical fibers, thus producing a good agreement with the results reported here.

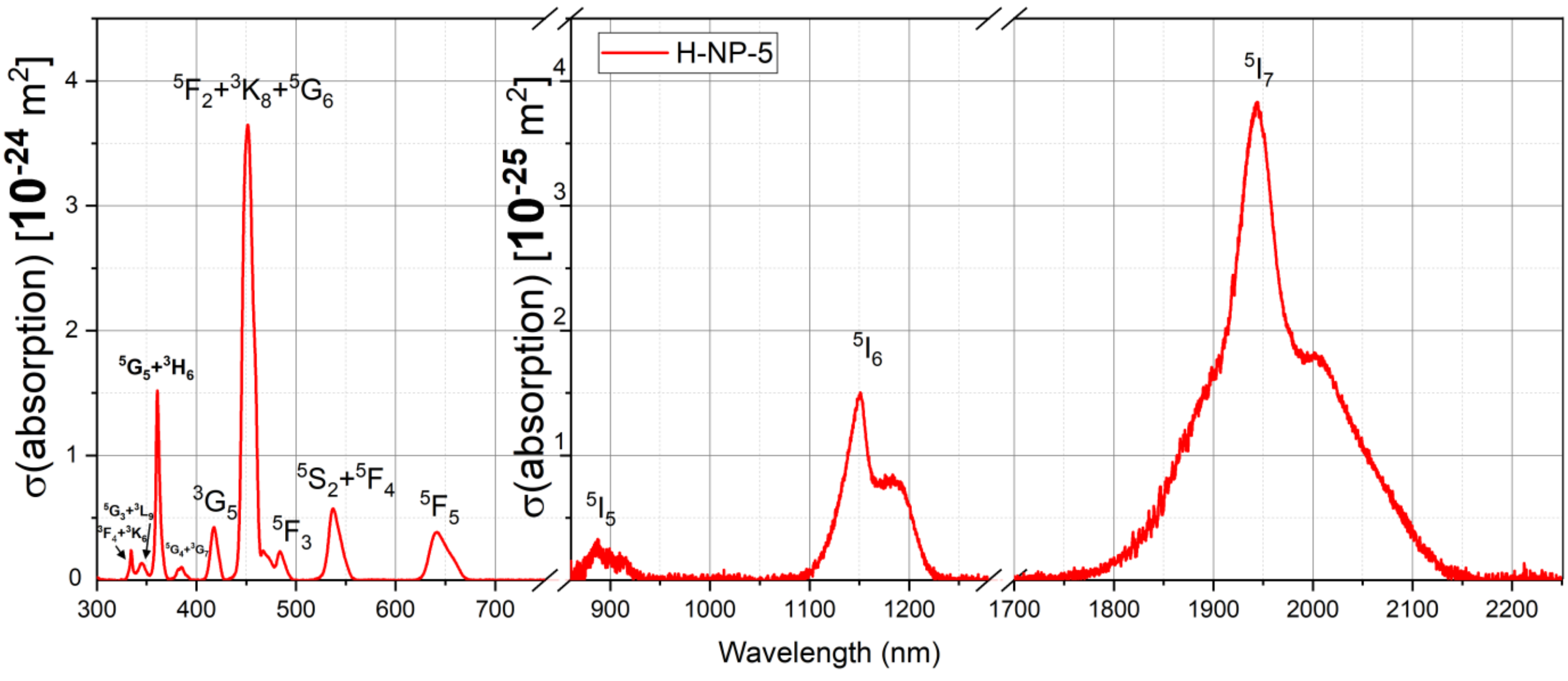

*Figure 2 – Spectroscopic properties of a representative preform H-NP-5, a) as-measured absorbance spectrum (red full line) and the background absorption baseline (black dotted line), b) absorption cross section spectrum, inset shows the weak $^5I_8$-$^5I_5$ band, c) detail of the absorption cross section of the $^5I_6$ and $^5I_7$ bands.*

On the contrary, the absorption bands corresponding to $^5I_8 \rightarrow {}^5G_6$ and $^5I_8 \rightarrow {}^3H_6$ at 450 nm and 360 nm, resp., are highly dependent on the concentration of $Al_2O_3$, while showing no specific correlation with Ho doping level. The difference between the spectra of two representative preforms with different $Al_2O_3$ content but nearly identical $Ho^{3+}$ concentration is depicted in Figure 4 for better visual comparison. The absorption cross section at 450 nm is approx. 4.5 $10^{-24}$ $m^2$ for the preforms containing up to 4 mol. %, regardless whether they were prepared by solution or nanoparticle doping and regardless of Ho doping level. On the contrary, the preforms containing 8 – 10 mol. % $Al_2O_3$ possess a significantly lower value of $\sigma_{ABS}$(450 nm) around 3.6 $10^{-24}$ $m^2$, again constant regardless of the holmium concentration. Similar effect can be observed for the $^3H_6$ (≈360 nm) absorption band. These results suggest that the $Ho^{3+}$ environment differs significantly depending on the content of $Al_2O_3$. The aforementioned transitions are known as hypersensitive, as they possess a large value of the $|U^2|$ reduced squared matrix element, order of magnitude larger compared to most other transitions [36]. The $|U^2|$ element is physically connected to the covalence of the RE-ligand bond and the symmetry of the crystal field [18] and the hypersensitive transitions are thus highly sensitive to even slight changes in the RE ion environment.

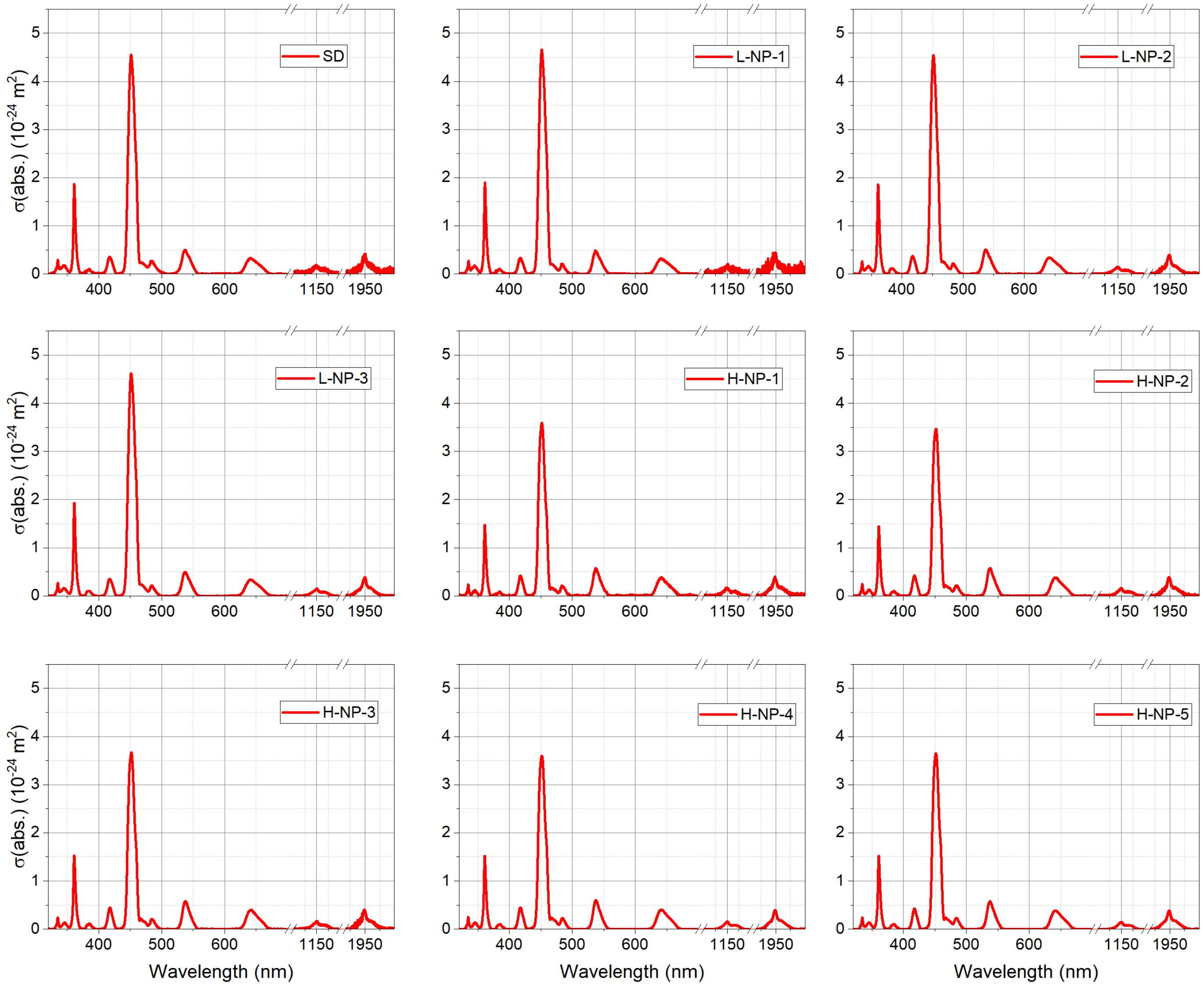

*Figure 3 – Absorption cross section spectra of all the investigated preforms.*

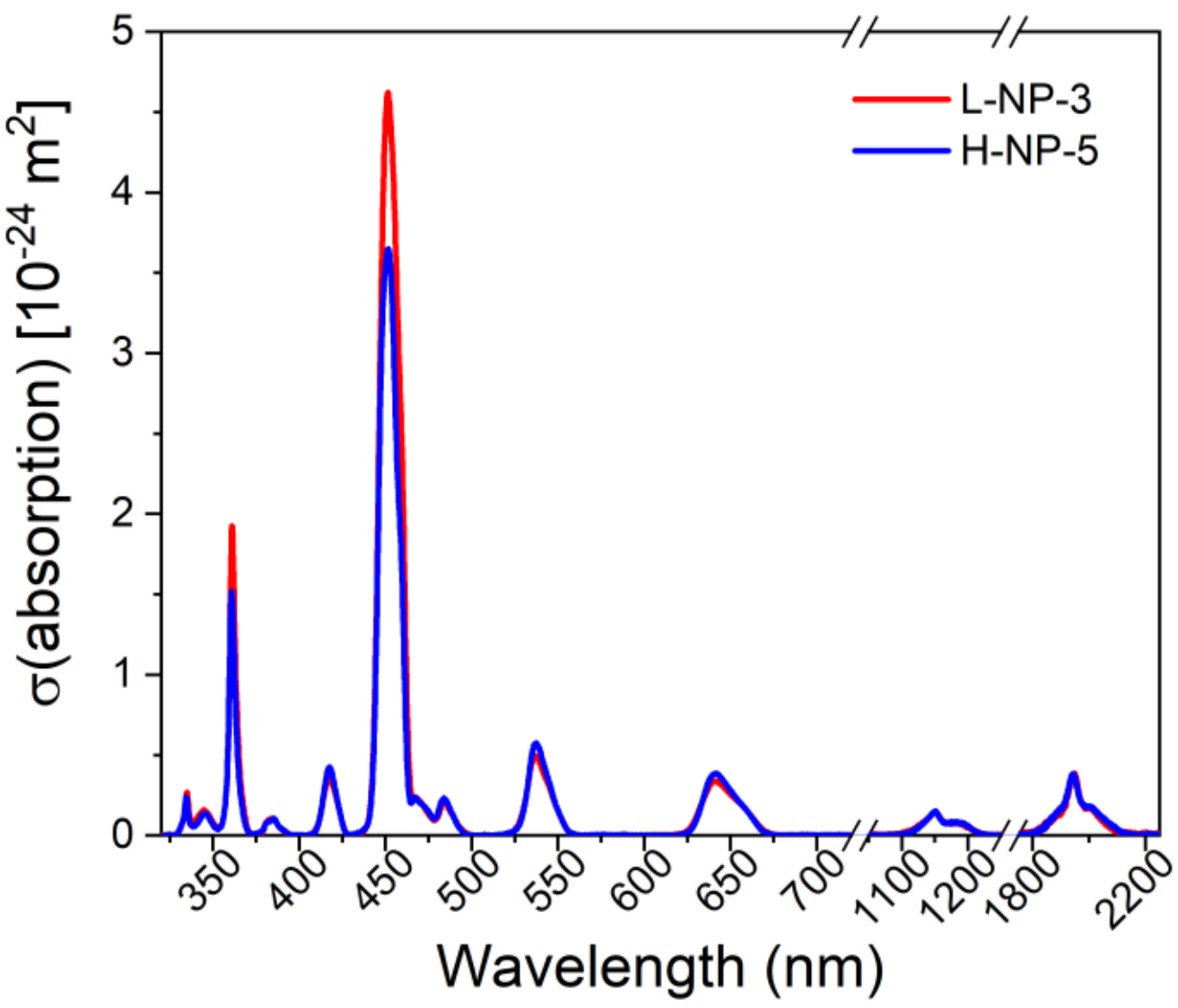

*Figure 4 – Comparison between the absorption cross section spectra of L-NP-3 and H-NP-5 preforms.*

### 3.3. *Judd-Ofelt analysis*

The absorption cross section bands corresponding to the individual transitions were integrated, the experimental line strengths, $S_{exp}$ were calculated and the JO analysis was carried out according to procedure described in [18]. Several things must be pointed out. First, the absorption band $^5I_8 \rightarrow {}^5I_5$ at 900 nm was not employed for the calculation, as it was only observable in the H-NP-4 and H-NP-5 preforms. We confirmed that this band has virtually no impact on the results of JO analysis, see Supplementary material S1. Second, $Ho^{3+}$ ion possesses a complex energy level structure and spectrum with overlapping or conjoined bands, which cannot be reliably separated or deconvoluted, these bands were thus integrated and used in the JO calculation together. The most notable example is the cluster of absorption bands around 450 nm, containing the hypersensitive $^5I_8 \rightarrow {}^5G_6$ transition, along with three less distinct bands.

The experimentally obtained values of line strengths, $S_{exp}$, along with the calculated values, $S_{cal}$, obtained by least-squares method from the JO analysis are summarized in Table 2 (for preforms up to 4 mol. % $Al_2O_3$) and Table 3 (for preforms 8 – 10 mol. % $Al_2O_3$). The experimental and calculated values are in reasonable agreement, the values of root-mean-square deviation, *σ(RMS)*, are below 0.32 $10^{-20}$ (≈ 1%), suggesting a good quality of the least-squares fit.

*Table 2 – the approximate values of mean wavelengths (λ), the experimental and calculated values of line strengths (S) and the root-mean-square deviation σ(RMS) values for the preforms containing up to 4 mol. % $Al_2O_3$. All line strength values are listed as $10^{-20}$ $cm^2$. Note that the values of mean wavelengths might differ by a few nanometers between the preforms, for exact values please see the LOMS software input files available at [32].*

| | | **SD** | | **L-NP-1** | | **L-NP-2** | | **L-NP-3** | |
|---|---|---|---|---|---|---|---|---|---|
| | **≈λ (nm)** | $S_{exp}$ | $S_{cal}$ | $S_{exp}$ | $S_{cal}$ | $S_{exp}$ | $S_{cal}$ | $S_{exp}$ | $S_{cal}$ |
| $^5I_7$ | 1960 | 2.06 | 2.43 | 2.16 | 2.47 | 2.08 | 2.35 | 1.95 | 2.26 |
| $^5I_6$ | 1155 | 1.26 | 1.02 | 1.36 | 1.04 | 0.92 | 0.98 | 0.97 | 0.94 |
| $^5F_5$ | 646 | 1.77 | 1.76 | 1.72 | 1.68 | 1.82 | 1.76 | 1.84 | 1.74 |
| $^5S_2$+$^5F_4$ | 539 | 2.08 | 1.73 | 1.88 | 1.71 | 2.05 | 1.69 | 1.98 | 1.64 |
| $^5F_{3+2}$+$^3K_8$+$^5G_6$ | 455 | 19.53 | 19.58 | 19.21 | 19.28 | 19.00 | 19.08 | 19.51 | 19.58 |
| $^3G_5$ | 418 | 1.16 | 1.35 | 1.06 | 1.21 | 1.14 | 1.39 | 1.13 | 1.41 |
| $^5G_4$+$^3G_7$ | 385 | 0.35 | 0.24 | 0.33 | 0.23 | 0.39 | 0.23 | 0.34 | 0.23 |
| $^5G_5$+$^3H_6$ | 362 | 3.34 | 2.98 | 3.41 | 2.92 | 3.52 | 2.92 | 3.50 | 2.99 |
| $^5G_3$+$^3L_9$ | 345 | 0.66 | 0.41 | 0.67 | 0.42 | 0.62 | 0.39 | 0.61 | 0.39 |
| $^3F_4$+$^3K_6$ | 334 | 0.53 | 0.36 | 0.41 | 0.32 | 0.40 | 0.36 | 0.55 | 0.37 |
| σ(RMS) | | 0.29 (<1 %) | | 0.29 (<1 %) | | 0.32 (1 %) | | 0.31 (<1 %) | |

*Table 3 – the approximate values of mean wavelengths (λ), the experimental and calculated values of line strengths (S) and the root-mean-square deviation σ(RMS) values for the preforms containing up to 8 – 10 mol. % $Al_2O_3$. All line strength and RMS deviation values are listed as $10^{-20}$ cm$^2$. Note that the values of mean wavelengths might differ by a few nanometers between the preforms, for exact values please see the LOMS software input files available at [32].*

| | ≈λ (nm) | H-NP-1 | | H-NP-2 | | H-NP-3 | | H-NP-4 | | H-NP-5 | |
|---|---|---|---|---|---|---|---|---|---|---|---|
| | | $S_{exp}$ | $S_{cal}$ | $S_{exp}$ | $S_{cal}$ | $S_{exp}$ | $S_{cal}$ | $S_{exp}$ | $S_{cal}$ | $S_{exp}$ | $S_{cal}$ |
| $^5I_7$ | 1960 | 2.01 | 2.29 | 1.91 | 2.28 | 2.02 | 2.34 | 2.12 | 2.38 | 2.07 | 2.36 |
| $^5I_6$ | 1155 | 1.02 | 0.96 | 1.01 | 0.96 | 1.06 | 0.98 | 0.97 | 1.00 | 0.94 | 0.99 |
| $^5F_5$ | 646 | 1.85 | 1.78 | 1.86 | 1.77 | 1.98 | 1.88 | 2.00 | 1.90 | 1.92 | 1.84 |
| $^5S_2$+$^5F_4$ | 539 | 2.06 | 1.71 | 2.09 | 1.70 | 2.11 | 1.77 | 2.14 | 1.80 | 2.16 | 1.76 |
| $^5F_{3+2}$+$^3K_8$+$^5G_6$ | 455 | 14.53 | 14.57 | 0.60 | 0.40 | 15.16 | 15.22 | 15.14 | 15.19 | 15.37 | 15.43 |
| $^3G_5$ | 418 | 1.22 | 1.41 | 14.55 | 14.59 | 1.30 | 1.53 | 1.31 | 1.54 | 1.24 | 1.47 |
| $^5G_4$+$^3G_7$ | 385 | 0.26 | 0.22 | 1.18 | 1.41 | 0.36 | 0.23 | 0.32 | 0.23 | 0.35 | 0.23 |
| $^5G_5$+$^3H_6$ | 362 | 2.64 | 2.29 | 0.31 | 0.22 | 2.86 | 2.40 | 2.78 | 2.40 | 2.86 | 2.42 |
| $^5G_3$+$^3L_9$ | 345 | 0.32 | 0.34 | 2.67 | 2.29 | 0.46 | 0.35 | 0.43 | 0.35 | 0.49 | 0.36 |
| $^3F_4$+$^3K_6$ | 334 | 0.31 | 0.36 | 0.41 | 0.34 | 0.37 | 0.39 | 0.36 | 0.39 | 0.37 | 0.38 |
| σ(RMS) | | 0.23 (<1 %) | | 0.26 (1 %) | | 0.27 (1 %) | | 0.24 (<1 %) | | 0.28 (1 %) | |

The results of the JO analysis are further summarized in Table 4, which lists the obtained JO parameters, $\Omega_2$, $\Omega_4$, and $\Omega_6$. Several observations can be made. The $\Omega_4$ and $\Omega_6$ parameters are relatively insensitive to the composition and show only a mild variance between the preforms. The only exception is L-NP-1 preform with a low $\Omega_4$ parameter of 2.3 $10^{-20}$ cm$^2$. However, it must be noted that the L-NP-1 contained the lowest concentration of $Ho^{3+}$ ions, and the absorption spectrum especially in the near-infrared region was relatively noisier compared to other samples. Nevertheless, the values of $\Omega_4$ and $\Omega_6$ parameters of all preforms fall into the error bounds of each other. The spectroscopic quality factor calculated as $\Omega_4/\Omega_6$ equals to approx. 2 in the case of all fabricated preforms, suggesting a high potential for efficient laser operation.

On the contrary, the value of $\Omega_2$ parameter is heavily dependent of the $Al_2O_3$ content, with value above 10.0 $10^{-20}$ cm$^2$ for preforms up to 4 mol. % $Al_2O_3$, and around 7.5 $10^{-20}$ cm$^2$ for preforms in 8 – 10 mol. % $Al_2O_3$ range. This observation is in a good agreement with the trend observed for the absorption cross section at 450 nm, since the $^5I_8 \rightarrow ^5G_6$ hypersensitive transition possesses a very large value of $|U^2|$ reduced squared matrix element and thus exerts significant impact on the $\Omega_2$ parameter.

*Table 4 – results of the JO analysis: The three JO parameters of the investigated preforms along with available values from literature.*

| | $\Omega_2$ ($10^{-20}$ cm$^2$) | $\Omega_4$ ($10^{-20}$ cm$^2$) | $\Omega_6$ ($10^{-20}$ cm$^2$) | Ref |
|---|---|---|---|---|
| *Solution doping* | | | | |
| SD | 10.6 ± 0.3 | 2.5 ± 0.4 | 1.2 ± 0.2 | - |
| *NP doping ($Al_2O_3$ < 4 mol. %)* | | | | |
| L-NP-1 | 10.5 ± 0.3 | 2.3 ± 0.4 | 1.3 ± 0.2 | - |
| L-NP-2 | 10.3 ± 0.3 | 2.6 ± 0.5 | 1.1 ± 0.2 | - |
| L-NP-3 | 10.6 ± 0.3 | 2.6 ± 0.5 | 1.1 ± 0.2 | - |
| *NP doping ($Al_2O_3$ ≈ 8 – 10 mol. %)* | | | | |
| H-NP-1 | 7.2 ± 0.2 | 2.7 ± 0.3 | 1.2 ± 0.1 | - |
| H-NP-2 | 7.3 ± 0.3 | 2.6 ± 0.4 | 1.2 ± 0.1 | - |
| H-NP-3 | 7.6 ± 0.3 | 2.9 ± 0.4 | 1.2 ± 0.2 | - |
| H-NP-4 | 7.6 ± 0.2 | 2.9 ± 0.4 | 1.2 ± 0.1 | - |
| H-NP-5 | 7.8 ± 0.3 | 2.8 ± 0.4 | 1.2 ± 0.1 | - |
| *$Al_2O_3$-$SiO_2$ prepared by sol-gel* | | | | |
| 4.5$Al_2O_3$-95.5$SiO_2$ | 8.83 | 2.45 | 1.06 | [19] |
| 30$Al_2O_3$-70$SiO_2$ | 3.056 | 2.346 | 1.955 | [20] |
| 4$Al_2O_3$-96$SiO_2$ | 4.69 | 2.90 | 0.93 | [21] |
| $Al_2O_3$-$SiO_2$ (unspecified) | 2.05 | 1.55 | 1.68 | [22] |
| *Other silicate glass* | | | | |
| $Na_2O$-$Al_2O_3$-$SiO_2$ | 6.01 | 1.53 | 1.03 | [37] |
| $Na_2O$-$GeO_2$-$SiO_2$ | 4.23 | 0.74 | 0.70 | [37] |
| $Na_2O$-ZnO-$SiO_2$ | 5.15 | 1.03 | 0.93 | [37] |
| $Li_2O$-$Na_2O$-SrO-$Al_2O_3$-$SiO_2$ | 3.60 | 2.30 | 0.65 | [23] |

### *3.4. Transition and emission analysis*

The JO parameters were subsequently employed to calculate the electric dipole transition probabilities, and along with magnetic dipole transition probabilities used to calculate the branching ratios and radiative lifetimes of all available transition up to $^3G_5$ excited level. The complete results of all 9 investigated samples are too voluminous to be included in the main body of paper and can be found in Supplementary material S2. These results may be freely used in calculations, simulations and modelling of HDFL devices or other related calculations, which require the values of absorption cross sections, transition probabilities, branching ratios or radiative lifetimes.

We employed the calculated values of radiative lifetimes of the $^5I_7 \rightarrow ^5I_8$ transition (1.9 – 2.2 µm emission), $\tau_{JO}$, and the experimentally measured values, $\tau_{EXP}$, to estimate the quantum efficiency, $\eta$, of the emission, according to Eq. (4).

$$\eta = \frac{\tau_{EXP}}{\tau_{JO}} \quad (4)$$

The experimentally measured decay curves of the 2 µm emission of all investigated preforms are depicted in Figure 5a for preforms up to 4 mol. % $Al_2O_3$ and Figure 5b for preforms containing 8 – 10 mol. % $Al_2O_3$ for clarity (employing different *x*-scale). The decay curves exhibited a single exponential character with only minor deviations and were reasonably well-fitted by single exponential function, allowing to extract the experimental fluorescence lifetime. The values of both experimental lifetime as well as the calculated one from JO analysis are summarized in Table 5.

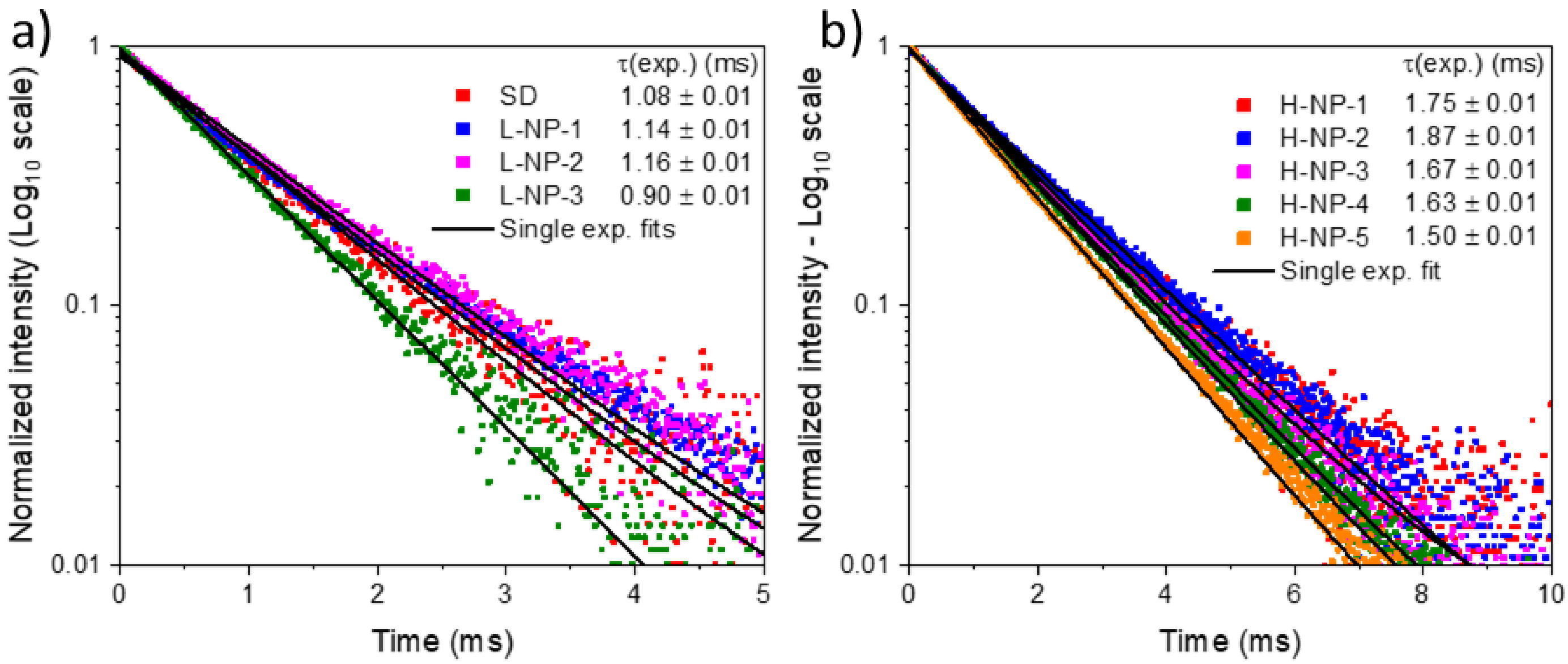

*Figure 5 – Fluorescence decay curves of the $^5I_7 \rightarrow {}^5I_8$ transition (2 µm emission), a) preforms from SD and NP-low groups, b) preforms from NP-high group.*

*Table 5 – Summarized values of measured, experimental lifetime, calculated lifetime from JO analysis and quantum efficiency of the $^5I_7 \rightarrow {}^5I_8$ transition (2 µm emission) in all investigated preforms and available literature.*

| | τ(exp.) (ms) | τ(JO) (ms) | η (%) | Ref. |
|---|---|---|---|---|
| *Solution doping* | | | | |
| SD | 1.08 | 16.3 | 6.6 | - |
| *NP doping ($Al_2O_3$ < 4 mol. %)* | | | | |
| L-NP-1 | 1.14 | 16.2 | 7.0 | - |
| L-NP-2 | 1.16 | 16.8 | 6.9 | - |
| L-NP-3 | 0.9 | 17.2 | 5.2 | - |
| *NP doping ($Al_2O_3$ ≈ 8 – 10 mol. %)* | | | | |
| H-NP-1 | 1.75 | 16.5 | 10.6 | - |
| H-NP-2 | 1.87 | 16.5 | 11.3 | - |
| H-NP-3 | 1.67 | 16.3 | 10.3 | - |
| H-NP-4 | 1.63 | 16.1 | 10.1 | - |
| H-NP-5 | 1.50 | 16.3 | 9.2 | - |
| *Other* | | | | |
| $Al_2O_3$-$SiO_2$ prepared by sol-gel | 1.124 | 13.89 | 8.09 | [19] |
| $Li_2O$-$Na_2O$-SrO-$Al_2O_3$-$SiO_2$ | 0.32 | 16.22 | 2.0 | [23] |

The measured values range between 0.9 – 1.9 ms, which is in a reasonable agreement with values previously obtained on optical fibers of similar compositions [26,27]. As obvious from the results, the preforms containing high content of $Al_2O_3$ in 8 – 10 mol. % range display significantly higher values of experimental lifetime, with up to 50 % increase compared to lower $Al_2O_3$ contents. The radiative lifetime of all samples regardless of $Al_2O_3$ or $Ho^{3+}$ content reaches values between 16.1 and 17.2 ms, thus producing quantum efficiency values in 5 – 7 % range for low $Al_2O_3$ preforms and 9 – 11 % for high $Al_2O_3$ preforms.

## 4. Discussion

The JO analysis provides a valuable insight into the structural characteristics of the material and the RE ion environment. The three JO parameters reported in this work follow the $\Omega_2 > \Omega_4 > \Omega_6$ trend, typical for silica-based glass matrices, and the values are generally in line with results reported on $Tm^{3+}$ and $Er^{3+}$ ions in alumino-silicate preforms of similar compositions [31,38]. The generally high values of $\Omega_2$ parameter above 7.0 $10^{-20}$ $cm^2$ are evidence of a relatively low symmetry of the $Ho^{3+}$ ion sites in the amorphous matrix as well as high covalency of the Ho-O bonds [12].

The JO and transition analysis allows us to evaluate the influence of the fabrication method and composition on the $Ho^{3+}$ ion environment. When comparing the preforms prepared by solution or nanoparticle doping, but with similar compositions, i.e., low content of $Al_2O_3$ up to 4 mol. % (SD and L-NP samples), it is evident that the results are nearly identical. These preforms exhibit nearly identical value of the JO parameters, experimental lifetime and quantum efficiency regardless of the preparation method. This result is another evidence that the $Al_2O_3$ nanoparticles used in the nanoparticle doping are fully dissolved during the preform sintering and collapse, with temperatures reaching up to 2100 °C, as was reported in our previous studies on $Ho^{3+}$ as well as $Tm^{3+}$-doped alumino-silicate preforms and fibers [26,27,39]. The $Ho^{3+}$ ion environment is thus identical in preforms prepared by the solution and nanoparticle doping, with the $Ho^{3+}$ ions likely embedded in a homogenous alumino-silicate amorphous matrix.

The main benefit of nanoparticle doping lies in the possibility of incorporating higher contents of $Al_2O_3$, which allows to also use higher doping of RE ions and achieve longer lifetimes and better efficiencies, as shown in our previous works [8,27,39] as well as here. When examining the results of the JO and transition analysis of the preforms containing higher $Al_2O_3$ contents some differences can be observed compared to the low-doped preforms. The $\Omega_2$ parameter exhibits a notably lower value around 7.5 $10^{-20}$ $cm^2$, which indicates a distinctly different $Ho^{3+}$ environment. The higher $Al_2O_3$ content also leads to a significantly longer measured lifetime, owing to the reduction of multiphonon relaxation, and a higher quantum efficiency. These observations are in line with our previous findings on both $Tm^{3+}$- and $Ho^{3+}$-doped optical fibers [27,39]. We previously reported that in alumino-silicate glass containing above 8 mol. % $A_2O_3$, a phase separation occurs, resulting in the creation of secondary $Al_2O_3$-enriched amorphous nanoparticles, which present a highly beneficial, low-phonon environment for the $Ho^{3+}$ ions [39,40]. The high $Al_2O_3$ concentrations, achievable by nanoparticle doping, thus lead to more favorable emission properties compared to the standard solution-doped fibers.

As stated in the introduction, despite the ubiquity of fiber lasers based on $Ho^{3+}$-doped alumino-silicate fibers, the available literature shows a surprising lack of relevant JO analysis results. A detailed study of $Ho^{3+}$-doped alumino-silicate bulk glass prepared by sol-gel method was presented by Wang et al. [19], with comparable results to our work, see Table 4 and Table 5. On the other hand, vastly different results were reported in works [20–22], with the most notable difference being the lower $\Omega_2$ parameter compared to our work here as well as Wang et al. The critical analysis of these discrepancies is difficult due to the lack of detail in the referenced works. First possible explanation is the fact that all these works only present absorption measurements in the 400 – 750 nm range, completely omitting the UV and NIR transitions. However, when we recalculated the JO analysis of the H-NP-5 preform omitting the same bands, the resulting parameters were $\Omega_2$ = 7.8 $10^{-20}$ $cm^2$, $\Omega_4$ = 2.3 $10^{-20}$ $cm^2$ and $\Omega_6$ = 1.7 $10^{-20}$ $cm^2$ (see Supplementary material S1 for more details). These values are relatively similar to those reported in Table 4 for full-range calculation, and the exclusion of the UV and NIR bands thus does not explain the discrepancies between our work and literature. A closer look at the measured data in [20–22], shows a significant difference in the intensity of hypersensitive transition $^5I_8 \rightarrow ^5G_6$ at 450 nm relative to other bands, when compared to our work or Wang et al. [19]. The $\Omega_2$ parameter is heavily dependent on the 450 nm transition, which possesses a very high value of $|U^2|$ matrix element, which thus explains the lower $\Omega_2$ parameter in the referenced works. However, the source of this discrepancy in absorption measurements is impossible to identify without further knowledge of experimental details and structural characterization in the referenced works. The lower measured absorption of the 450 nm transition may be caused by both experimental effects, e.g., parasitic emission excited by the incident light or detector over-saturation, as well as differences in the $Ho^{3+}$ environment, as the samples in [20–22] were prepared by sol-gel method and sintered only up to 1000 °C, whereas the preforms in our work were collapsed at temperatures up to 2100 °C.

Of special interest is the comparison of our results with the JO analysis of multicomponent lithium-sodium-strontium-alumino silicate glass by Peng et al. [23], which is used as a source of JO and transition parameters in several high-profile works dealing with HDFL based on alumino-silicate fibers [15,17]. As apparent from the comparison in Table 6, the JO parameters differ significantly, which may potentially lead to unwanted errors and deviation in the simulation. Surprisingly, however, the radiative lifetimes, $\tau_{JO}$, and branching ratios, $\beta$, for the first three excited levels are relatively similar, rendering the published simulations generally reliable. Still, for any calculation or simulation, it is advised to use data from as close composition as possible for maximum accuracy, as the radiative parameters may differ significantly with composition.

*Table 6 – comparison between the JO parameters ($\Omega_{2,4,6}$), radiative lifetimes ($\tau_{JO}$) and branching ratios (β) for the first three excited levels for H-NP-5 sample and multicomponent silicate glass by Peng et al.*

| | H-NP-5 | Peng et al. [23] |
|---|---|---|
| $\Omega_2$ ($10^{-20}$ cm$^2$) | 7.7 | 3.60 |
| $\Omega_4$ ($10^{-20}$ cm$^2$) | 2.7 | 2.30 |
| $\Omega_6$ ($10^{-20}$ cm$^2$) | 1.2 | 0.65 |
| $\tau_{JO}(^5I_7 \rightarrow {}^5I_8)$ (ms) | 16.3 | 16.22 |
| $\tau_{JO}(^5I_6 \rightarrow {}^5I_8)$ (ms) | 8.3 | 11.4 |
| $\tau_{JO}(^5I_5 \rightarrow {}^5I_8)$ (ms) | 11 | 14.8 |
| $\beta(^5I_6 \rightarrow {}^5I_8)$ (-) | 0.822 | 0.828 |
| $\beta(^5I_6 \rightarrow {}^5I_7)$ (-) | 0.178 | 0.172 |
| $\beta(^5I_5 \rightarrow {}^5I_6)$ (-) | 0.393 | 0.416 |
| $\beta(^5I_5 \rightarrow {}^5I_7)$ (-) | 0.509 | 0.444 |
| $\beta(^5I_5 \rightarrow {}^5I_8)$ (-) | 0.098 | 0.140 |

## 5. Conclusion

We present a detailed Judd-Ofelt analysis of $Ho^{3+}$-doped alumino-silicate optical fiber preforms in a wide range of compositions and prepared using both standard solution doping method as well as the modified nanoparticle doping. The absorption spectra were measured, the absorption cross sections were calculated for each observed transition and the JO analysis was conducted. The preforms exhibited the values of JO parameters of $\Omega_2$ = (7.2 – 10.6) $10^{-20}$ cm$^2$, depending on the $Al_2O_3$ content, and $\Omega_4$ and $\Omega_6$ around 2.5 and 1.2 $10^{-20}$ cm$^2$, respectively. The radiative transition parameters, such as transition probabilities, branching ratios and radiative lifetimes, were calculated. The radiative lifetime of the $^5I_7 \rightarrow {}^5I_8$ transition, combined with a measured value, was used to calculate the quantum efficiency. The preforms prepared by the nanoparticle doping containing high $Al_2O_3$ contents above 8 mol. % exhibited superior values of measured lifetime and quantum efficiency compared to the standard solution-doped samples. To the best of our knowledge, this work is the first comprehensive report on the JO analysis of $Ho^{3+}$-doped alumino-silicate glass, the calculated parameters may be freely used in various calculations, simulations and modelling of HDFL and other devices.

### Data availability statement

The data underlying the results of this manuscript are available in ref. [32].

### Declaration of competing interests

The authors declare that they have no known competing financial interests or personal relationships that could have appeared to influence the work reported in this paper.

### Acknowledgements

This work was supported by the Czech Science Foundation, grant No. GA26-22288S and grant No. GA26-21543S. This work was co-funded by the European Union and state budget of the Czech Republic under the project LasApp CZ.02.01.01/00/22_008/0004573.

**Credit authorship contribution statement**

**Petr Vařák:** Writing – review & editing, Writing – original draft, Methodology, Investigation, Formal analysis, Data curation, Conceptualization, Funding Acquisition. **Michal Kamrádek:** Writing – review & editing, Investigation. **Pavla Nekvindová:** Writing – review & editing, Supervision, Resources, Investigation. **Jan Hrabovský:** Writing – review & editing, Software. **Pavel Peterka:** Writing – review & editing, Supervision, Funding acquisition, Conceptualization.